\documentclass[preprint,eqsecnum,nofootinbib,natbib,aps]{revtex4}
\usepackage{graphicx}

\begin{document}
\title{Nonlinear Evolutions and Non-Gaussianity in Generalized Gravity}
\author{Seoktae Koh}
\email{kohst@ihanyang.ac.kr}
\author{Sang Pyo Kim}
\email{sangkim@kunsan.ac.kr}
\affiliation{Department of Physics, Kunsan National University, Kunsan,
573-701,Korea}
\author{Doo Jong Song}
\email{djsong@kao.re.kr}
\affiliation{ Korea Astronomy Observatory, Daejeon 305-348, Korea}
\begin{abstract}
We use the Hamilton-Jacobi theory to study the nonlinear
evolutions of inhomogeneous spacetimes during inflation in
generalized gravity. We find the exact solutions to the lowest
order Hamilton-Jacobi equation for special scalar potentials and
introduce an approximation method for general potentials. The
conserved quantity invariant under a change of timelike
hypersurfaces proves useful in dealing with gravitational
perturbations. In the long-wavelength approximation, we find a
conserved quantity related to the new canonical variable that
makes the Hamiltonian density vanish, and calculate the
non-Gaussianity in generalized gravity. The slow-roll inflation
models with a single scalar field in generalized gravity predict
too small non-Gaussianity to be detected by future CMB
experiments.
\end{abstract}

\maketitle

\section{Introduction}

Inflation scenario is a successful model to solve the problems of
the standard Big Bang theory and explains remarkably the
observational data. Quantum fluctuations of a scalar field are
adiabatic and Gaussian during the inflation period and provide a
seed for density perturbations. The amplitudes of the
perturbations freeze out when the perturbations stretch out to the
superhorizon scale by an accelerated expansion. Inflation also
gives the scale invariant spectrum ($n_s=1$) when the perturbation
modes cross the horizon. Linear perturbation theory is enough to
explain these gravitational perturbations and temperature
anisotropy in the early universe.

Recent WMAP observations \cite{komatsu03} try to find a signal of
the non-Gaussianity in the temperature anisotropy. The
non-Gaussian signal in the CMB anisotropy might be generated
either from the non-vacuum initial state \cite{gangui02}
 or from
nonlinear gravitational perturbation \cite{acquaviva03}. Gaussian
statistical properties are completely specified by the two-point
correlation function. However, the two-point correlation function
is not sufficient to describe the statistical properties of the
non-Gaussianity, so it is necessary to investigate higher order
correlations such as the three-point correlation for nonlinearity
of a perturbation field  or the four-point
correlation for a non-vacuum initial state. Second
order perturbation theory has been used to explain the
non-Gaussianity in the temperature anisotropy \cite{acquaviva03}. 
In addition to
second order perturbation theory, the Hamiltonian formalism turned
out to useful to deal with nonlinear evolutions in the early
universe and was applied to canonical quantum gravity
\cite{dewitt67} or semiclassical gravity \cite{banks85,kim92} for a long
time.

Salopek and Bond in Ref. \cite{salopek90} employed the Hamiltonian
formalism to study the nonlinear evolutions of gravitational
perturbations. It was also applied to Brans-Dicke theory
\cite{soda95} and low energy effective string theory
\cite{saygili99}. Especially, the Hamilton-Jacobi theory provides
a powerful tool to get solutions of nonlinear evolutions in the
early universe through a generating functional which satisfies the
momentum constraint equation. Even though it is difficult to get
exact solutions of the Hamilton-Jacobi equation for general
potentials, the large scale perturbation, which contributes mainly
to the large scale structure in the present Universe, could be
treated appropriately using the long wavelength approximation. The
long wavelength approximation assumes that the length scale of the
spatial variation is much longer than the Hubble radius, so it is
a reasonable assumption to deal with super-horizon scale
perturbations. It is also known that the gauge invariant conserved
quantity exists in nonlinear perturbation theory for a
superhorizon scale \cite{salopek90,rigopoulos03,lyth04}.

Brans-Dicke type gravity naturally emerges from the fundamental
theory of particle physics such as string or M-theory. Although it
is not clear how scalar fields couple to gravity, it is necessary
to investigate the perturbations in the alternative gravity theory
such as $f(\phi)R$ type gravity as well as in Einstein gravity.
Recent supernovae observation \cite{perlmutter99} and WMAP results
\cite{spergel03} imply that our universe today is in an
accelerated expansion phase and dominated by the dark energy which
has the equation of state, $p/\rho <-1/3$. It is also needed to
consider nonlinear evolutions of such a matter component to see
whether their existence affects the temperature anisotropy.

In this paper, the Hamilton-Jacobi formalism will be used to study
the nonlinear evolutions of inhomogeneous spacetimes in
generalized gravity theory during the inflation period. The
canonical variables will be transformed to new ones that make the
new Hamiltonian density vanish and these new variables are
constant in time for fixed spacelike hypersurfaces. The gauge
invariant quantity, $\zeta$, which is conserved in the large scale
limit, is derived from one of new canonical variables. By
introducing the nonlinear parameter, $f_{NL}$, the conserved
quantity may be decomposed into a linear Gaussian part, $\zeta_L$,
and a nonlinear part \cite{komatsu01}:
\begin{eqnarray}
\zeta({\bf x}) = \zeta_L({\bf x})+\frac{3}{5}f_{NL} \bigl[
\zeta_L^2({\bf x}) -\langle \zeta_L^2({\bf x})\rangle \bigr].
\end{eqnarray}
 Non-Gaussianity is parameterized through $f_{NL}$ which
will be constrained by observations. It is generally expected to be 
difficult
to detect a non-Gaussian signal in CMB experiments for a single
field inflation model in Einstein gravity \cite{acquaviva03}. So
the detection of the non-Gaussianity can constrain different
inflation models.

This paper is organized as follows. In Sec. II, we derive the
Hamilton and momentum constraint equations in generalized gravity.
The Hamilton-Jacobi equation will be obtained through a canonical
transformation. Assuming a generating functional which satisfies
the momentum constraint equation, we get the conserved quantity
that is invariant under a change of timelike hypersurfaces in the
long-wavelength approximation.  In Sec. III,  non-Gaussianity will
be computed using the generalized curvature perturbation on
comoving hypersurfaces. Finally, we discuss the physical
implications of the non-Gaussianity in generalized gravity in Sec.
IV.

\section{Hamilton-Jacobi Formalism in Generalized Gravity}

\subsection{Hamilton equations}

The generalized gravity action to be studied in this paper is
given by
\begin{eqnarray}
{\mathcal I} =\int \sqrt{-g}\Biggl[\frac{1}{2}f(\phi)R-\frac{1}{2}
\omega(\phi)g^{\mu\nu}\partial_{\mu}\phi \partial_{\nu}\phi
-V(\phi)\Biggr],
\end{eqnarray}
where $f(\phi),\omega(\phi)$ and $V(\phi)$ are functions of a
scalar field $\phi$. We shall confine our attention to the
slow-roll inflation models that are described by this action.
Einstein gravity is recovered when $f(\phi) = 1/8\pi G$ and
$\omega(\phi) = 1$. Further, Brans-Dicke theory is prescribed by
$f(\phi) = \phi/ 8\pi,~ \omega(\phi) = \omega/8 \pi \phi$ and
$V=0$; the non-minimally coupled scalar field theory corresponds
to $f(\phi)=(1 -8\pi G\xi \phi^2)/8\pi G$ and
$\omega(\phi)=1$; the low energy effective string theory is given
by $f(\phi)=e^{-\phi}$ and $\omega(\phi)=e^{-\phi}$. We consider
the Arnowitt-Deser-Misner (ADM) metric
\begin{eqnarray}
ds^2 = (-N^2+N_i N^i)dt^2 + 2N_i dt dx^i + \gamma_{ij}dx^i dx^j,
\end{eqnarray}
where $N$ and $N^i$ are a lapse function and a shift vector,
respectively, and $\gamma_{ij}$ is a 3-spatial metric. The
$4$-dimensional Ricci scalar, $R$, can be written in terms of the
3-dimensional Ricci scalar, ${}^3R$, and the extrinsic curvature,
$K_{ij}$, as \cite{mtw73}
\begin{eqnarray}
R = {}^3R+K_{ij}K^{ij}-K^2-\frac{2}{N\gamma^{1/2}}
\Biggl[\frac{\partial}{\partial t}
(\gamma^{1/2}K)-(\gamma^{1/2}KN^i-\gamma^{1/2}\gamma^{ij}N_{,j})_{,i}\Biggr].
\end{eqnarray}
The terms in the square bracket in the above formula are total
derivatives or surface terms so that they can be integrated out in
Einstein gravity.  They cannot, however, be neglected in
generalized gravity. The extrinsic curvature tensor and trace are
given by
\begin{eqnarray}
K_{ij}=\frac{1}{2N}\Biggl(N_{i|j}+N_{j|i}-\frac{\partial
\gamma_{ij}} {\partial t}\Biggr), \quad K=\gamma^{ij}K_{ij} =
K^i_i,
\end{eqnarray}
where the vertical bar is a covariant derivative with respect to
$\gamma_{ij}$. The $K$ is a generalization of the Hubble
parameter. By varying action with respect to $\dot{\gamma_{ij}}$
and $\dot{\phi}$, the momenta conjugate to $\gamma_{ij}$ and
$\phi$ are obtained as
\begin{eqnarray}
\pi^{ij} &=& -\frac{1}{2}\gamma^{1/2}f(K^{ij}-K\gamma^{ij})
-\frac{f_{,\phi}}{2N}\gamma^{1/2}\gamma^{ij}(\dot{\phi}-N^k\partial_k
\phi),
\\
\pi^{\phi} &=& f_{,\phi}\gamma^{1/2}K+\gamma^{1/2}\frac{\omega}{N}
(\dot{\phi}-N^i\partial_i \phi).
\end{eqnarray}
From these relations, $K$ can be written as
\begin{eqnarray}
K=\frac{1}{(1+3\Omega)\gamma^{1/2}}\Biggl[\frac{1}{f}\pi^{\gamma}
+\frac{3\Omega}{f_{,\phi}}\pi^{\phi}\Biggr],
\end{eqnarray}
where $\pi^{\gamma}=\gamma^{ij}\pi_{ij}$. Variations of the action
with respect to $N$ and $N^i$ lead to the Hamiltonian and momentum
constraint equations, respectively,
\begin{eqnarray}
\mathcal{H} &=& \frac{2}{\gamma^{1/2}f}\pi^{ij}\pi^{kl}
\Biggl[\gamma_{ik}\gamma_{jl}-\frac{1+2\Omega}{2(1+3\Omega)}\gamma_{ij}
\gamma_{kl}\Biggr]+\frac{1}{2(1+3\Omega)}\frac{1}{\gamma^{1/2}\omega}
(\pi^{\phi})^2                   \nonumber \\
& & -\frac{2\Omega}{1+3\Omega}\frac{1}{\gamma^{1/2}f_{,\phi}}
\pi^{\gamma}\pi^{\phi}-\frac{1}{2}\gamma^{1/2}f ~{}^{3}R +
\frac{1}{2}\gamma^{1/2}(2 f_{,\phi\phi}+\omega)\gamma^{ij}
\partial_i \phi \partial_j \phi             \nonumber \\
& & +\gamma^{1/2}V+\gamma^{1/2}f_{,\phi}
\Delta \phi =0, \label{hamilton}\\
\mathcal{H}_i &=& -2\partial_j (\gamma_{ik}\pi^{kj})
+\pi^{kl}\partial_i \gamma_{kl}+\pi^{\phi}\partial_i \phi =0,
\label{momentum}
\end{eqnarray}
where
\begin{eqnarray}
\Omega (\phi) \equiv \frac{f_{,\phi}^2}{2f \omega},\quad \Delta
\phi ={\phi^{|i}}_{|i}.
\end{eqnarray}
Then one finds the action of the form
\begin{eqnarray}
\mathcal{I}=\int d^4 x [\pi^{ij}\dot{\gamma}_{ij}+\pi^{\phi}
\dot{\phi}-N\mathcal{H}-N^i \mathcal{H}_i]. \label{action2}
\end{eqnarray}
Here, $N$ and $N^i$ are considered as Lagrange multipliers. The
evolution equations for $\gamma_{ij}$ and $\phi$ can be obtained
by varying the action (\ref{action2}) with respect to $\pi^{ij}$
and $\pi^{\phi}$:
\begin{eqnarray}
& &\frac{1}{N}(\dot{\gamma}_{ij}-N_{i|j}-N_{j|i})=
\frac{4}{\gamma^{1/2}f}\Biggl[\pi_{ij}-\frac{1+2\Omega}{2(1+3\Omega)}
\pi^{\gamma}\gamma_{ij}\Biggr]-\frac{2\Omega}{1+3\Omega}\frac{1}{\gamma^{1/2}
f_{,\phi}}\gamma_{ij}\pi^{\phi}, \\
& &\frac{1}{N}(\dot{\phi}-N^i\partial_i \phi)
=\frac{1}{(1+3\Omega)\gamma^{1/2}}
\Biggl[\frac{1}{\omega}\pi^{\phi}-\frac{2\Omega}{f_{,\phi}}\pi^{\gamma}
\Biggr].
\end{eqnarray}

\subsection{Hamilton-Jacobi equation}

To solve the Hamiltonian and the momentum constraint equations,
(\ref{hamilton}) and (\ref{momentum}), we use the Hamilton-Jacobi
theory. Through an appropriate canonical transformation from
$\gamma_{ij},\phi, \pi^{ij}$ and $\pi^{\phi}$ to new ones
$\tilde{\gamma}_{ij},\tilde{\phi}, \tilde{\pi}^{ij}$ and
$\tilde{\pi}^{\phi}$, we can construct a vanishing Hamiltonian
\begin{eqnarray}
\tilde{H} = H+\frac{\partial S}{\partial t} =0,
\end{eqnarray}
where
\begin{eqnarray}
H=\int d^3 x (N \mathcal{H}+N^i \mathcal{H}_i),
\end{eqnarray}
and the generating functional $S$ is a function of
$\phi,\gamma_{ij},\tilde{\phi}$ and $\tilde{\gamma}_{ij}$. Then
the canonical transformation gives the following relations
\begin{eqnarray}
\pi^{ij} =\frac{\delta S}{\delta \gamma_{ij}},\quad
\pi^{\phi}=\frac{\delta S}{\delta \phi}, \quad \tilde{\pi}^{ij}
=\frac{\delta S}{\delta \tilde{\gamma}_{ij}},\quad
\tilde{\pi}^{\phi}= \frac{\delta S}{\delta \tilde{\phi}}.
\label{canonical}
\end{eqnarray}
Finally, we get the Hamilton-Jacobi equation from the Hamiltonian
constraint
\begin{eqnarray}
& &\frac{2}{\gamma^{1/2}f}\frac{\delta S}{\delta \gamma_{ij}}
\frac{\delta S}{\delta \gamma_{kl}}\Biggl[\gamma_{ik}\gamma_{jl}
-\frac{1+2\Omega}{2(1+3\Omega)}\gamma_{ij}\gamma_{kl}\Biggr]
+\frac{1}{2(1+3\Omega)}\frac{1}{\gamma^{1/2}\omega}
\Biggl(\frac{\delta S}{\delta \phi}\Biggr)^2  \nonumber \\
& &
-\frac{2\Omega}{1+3\Omega}\frac{1}{\gamma^{1/2}f_{,\phi}}\gamma_{ij}
\frac{\delta S}{\delta \gamma_{ij}}\frac{\delta S}{\delta \phi}
-\frac{1}{2}\gamma^{1/2}f~{}^{3}R+\frac{1}{2}\gamma^{1/2}(2f_{,\phi\phi}
+\omega)\gamma^{ij}\partial_i \phi \partial_j \phi \nonumber \\
& & +\gamma^{1/2}V+\gamma^{1/2}f_{,\phi}\Delta \phi =0,
\label{hjeq}
\end{eqnarray}
and the momentum constraint equation
\begin{eqnarray}
-2\partial_j \Biggl(\gamma_{ik}\frac{\delta S}{\delta
\gamma_{kj}}\Biggr)+\frac{\delta S}{\delta \gamma_{kl}}\partial_i
\gamma_{kl} +\frac{\delta S}{\delta \phi}\partial_i \phi=0.
\end{eqnarray}
The momentum constraint equation implies that the generating
functional $S(\phi,\gamma_{ij},\tilde{\phi},\tilde{\gamma}_{ij})$
is invariant under spatial coordinate transformations. In general,
the momentum constraint equation does not vanish through canonical
transformations as long as $N^i$ does not vanish. On the contrary,
the Hamiltonian constraint vanishes strongly. It is difficult to
solve the Hamilton-Jacobi equation, (\ref{hjeq}), in general,
except for special cases such as an exponential potential in
Einstein gravity \cite{salopek90}.

\subsection{long-wavelength approximation}

To deal with the large scale gravitational perturbations, it is
reasonable to use the approximation that temporal variations of
fields are much greater than spatial variations. In inflation
scenario, the inhomogeneous field, whose physical wavelength is
much larger than the horizon size at the end of the inflation
period, mostly contributes to formation of the large scale
structure in the present universe and large angle CMB anisotropy.
The long wavelength approximation assumes that the characteristic
scale, $\lambda$, of spatial variations is much longer than the
Hubble radius, $H^{-1}$, \cite{salopek90,soda95}:
\begin{eqnarray}
\frac{1}{a}\partial_i \gamma_{jk} \ll \dot{\gamma}_{jk} \rightarrow
\lambda_{ph}=a\lambda \gg H^{-1},
\end{eqnarray}
where $H$ is a Hubble parameter, and $\lambda_{ph}$ and  $\lambda$
are a physical and a comoving wavelength, respectively. The
generating functional can be expanded in a series of spatial
gradient terms
\begin{eqnarray}
S = S^{(0)}+S^{(2)}+S^{(4)}+\cdots.
\end{eqnarray}
The lowest order Hamilton-Jacobi equation neglects the terms
containing spatial gradients. In this paper we only consider the
lowest order Hamilton-Jacobi equation which is sufficient for
dealing with the nonlinear evolution of the inhomogeneous
gravitational fields. Then the lowest order Hamilton-Jacobi
equation is
\begin{eqnarray}
& &\frac{2}{\gamma^{1/2}f}\frac{\delta S^{(0)}}{\delta
\gamma_{ij}} \frac{\delta
S^{(0)}}{\delta_{kl}}\Biggl[\gamma_{ik}\gamma_{jl}-
\frac{1+2\Omega}{2(1+3\Omega)}\gamma_{ij}\gamma_{kl}\Biggr]
+\frac{1}{2(1+3\Omega)}\frac{1}{\gamma^{1/2}\omega}
\Biggl(\frac{\delta S^{(0)}}{\delta \phi}\Biggr)^2 \nonumber \\
& &
-\frac{2\Omega}{1+3\Omega}\frac{1}{\gamma^{1/2}f_{,\phi}}\gamma_{ij}
\frac{\delta S^{(0)}}{\delta \gamma_{ij}}\frac{\delta
S^{(0)}}{\delta \phi} +\gamma^{1/2} V =0.
\end{eqnarray}
We assume that the lowest order generating functional takes the
following form
\begin{eqnarray}
S^{(0)}  (\phi, \gamma_{ij}, \theta) = 2\beta \int d^3 x
\gamma^{1/2}f^{3/2}(\phi) W(\phi,\gamma_{ij},\theta), \label{gen0}
\end{eqnarray}
where $\theta$ is a new canonical variable involving
$\tilde{\gamma}_{ij}$ and $\tilde{\phi}$, and $\beta$ is a
constant which carries a dimension. For Einstein gravity, $W$ can
be interpreted as a locally defined Hubble parameter if we take
$\beta = -\sqrt{8\pi G}$~ \cite{salopek90}. This generating
functional automatically satisfies the momentum constraint
equations if $N^i=0$, and this will be discussed in the next
section.

It is convenient to factor the 3-spatial metric $\gamma_{ij}$ into
a conformal factor and a conformal 3-spatial metric $h_{ij}$ with
the unit determinant ${\rm det} (h_{ij}) = 1$:
\begin{eqnarray}
\gamma_{ij}(t,{\bf x}) = \gamma^{1/3}(t,{\bf x}) h_{ij}({\bf x}).
\end{eqnarray}
The gravitational waves are related to the $h_{ij}$. Then, from
Eq.~(\ref{canonical}), the conjugate momenta for $\gamma_{ij}$ and
$\phi$ are
\begin{eqnarray}
\pi^{ij} &=& 2\beta f^{3/2}\gamma^{1/2}\Biggl[
\frac{1}{2}\gamma^{ij}W+\gamma^{-1/3}\Biggl\{\frac{\partial
W}{\partial h_{ij}}
-\frac{1}{3}h^{ij}h_{kl}\frac{\partial W}{\partial h_{kl}}\Biggr\}\Biggr], \\
\pi^{\phi} &=& 2\beta \gamma^{1/2}\Biggl[
\frac{3}{2}f^{1/2}\frac{\partial f}{\partial \phi}
+f^{3/2}\frac{\partial W}{\partial \phi}\Biggr],
\end{eqnarray}
where we have used \cite{salopek91}
\begin{eqnarray}
\frac{\partial W}{\partial \gamma_{ij}}
=\gamma^{-1/3}\Biggl[\frac{\partial W}{\partial h_{ij}}
-\frac{1}{3}h^{ij}h_{kl}\frac{\partial W}{\partial h_{kl}}\Biggr],
\end{eqnarray}
which is traceless, $\gamma_{ij} \partial W/\partial
\gamma_{ij}=0$. If we decompose $\pi^{ij}$ into the trace
$\pi^{\gamma}$ and the traceless part $\bar{\pi}^{ij}$, they are
given by
\begin{eqnarray}
\pi^{\gamma}=3\beta f^{3/2} \gamma^{1/2}W,\quad
\bar{\pi}^{ij}=2\beta f^{3/2} \gamma^{1/6} \Biggl[\frac{\partial
W}{\partial h_{ij}} -\frac{1}{3}h^{ij}h_{kl}\frac{\partial
W}{\partial h_{kl}}\Biggr].
\end{eqnarray}
As long as we do not concern about the gravitational radiations,
$W$ is assumed to be independent of $h_{ij}$, so $\bar{\pi}^{ij}$
can be set to zero. Then the Hamilton-Jacobi
equation becomes
\begin{eqnarray}
W^2 -\frac{2f}{3\omega(1+3\Omega)}\Biggl(\frac{\partial W}
{\partial \phi}\Biggr)^2-\frac{1}{3\beta^2 f^2}V=0, \label{hj}
\end{eqnarray}
and the evolution equations for $\gamma$ and $\phi$ are
\begin{eqnarray}
K &\equiv& -\frac{1}{2N}\frac{\dot{\gamma}}{\gamma} =\frac{3\beta
f^{1/2}}{1+3\Omega}\Biggl[(1+3\Omega)W
+\frac{f_{,\phi}}{\omega}\frac{\partial W}{\partial \phi}\Biggr],
\label {extrinsic}\\
\frac{1}{N}\dot{\phi} &=& \frac{2\beta f^{3/2}}{\omega(1+3\Omega)}
\frac{\partial W}{\partial \phi}. \label{scalar}
\end{eqnarray}

\subsection{Solutions of Hamilton-Jacobi equation}

Although the Hamilton-Jacobi equation (\ref{hj}) is difficult to
be exactly solved for general potentials, it can be solved for
some special cases in generalized gravity.

\subsubsection{$V=3\beta^2 f^2$}

If the potential is given by $V (\phi) =3\beta^2 f^2(\phi)$, then
the Hamilton-Jacobi equation takes the form
\begin{eqnarray}
\Biggl(\frac{\partial W}{\partial \phi}\Biggr)^2 -\alpha^2 W^2 +
\alpha^2 =0, \label{sol 1}
\end{eqnarray}
where
\begin{eqnarray}
\alpha^2(\phi) \equiv \frac{3\omega (1+3\Omega )}{2f}.
\end{eqnarray}
As $\alpha^2 > 0$,  Eq. (\ref{sol 1}) can be exactly integrated to
yield \cite{handbook}
\begin{eqnarray}
W (\phi) = \cosh[\pm \int \alpha (\phi) d\phi +C], \label{sol w}
\end{eqnarray}
where $C$ is an integration constant. The solution of the form
(\ref{sol w}) is available for a constant potential in Einstein
gravity or a massive scalar field potential for the generalized
gravity with $f(\phi) \propto \phi$.

\subsubsection{$V=0$ or general potentials} \label{homosect}

With the identification of  $W = q$ and $\phi = t$, the
Hamilton-Jacobi equation may be interpreted as a time-dependent
inverted oscillator with a unit mass, frequency $\alpha$, and
energy $\alpha^2 \rho$:
\begin{eqnarray}
\frac{1}{2} \Biggl(\frac{\partial W}{\partial \phi} \Biggr)^2 -
\frac{1}{2}\alpha^2 W^2 = - \frac{1}{2} \alpha^2 \rho,
\label{hjoscil}
\end{eqnarray}
where $\rho(\phi)=V/(3\beta^2 f^2)$. In fact, this inverted
oscillator has a $\phi$-dependent energy and curvature (spring
constant) of potential. The $W=\pm \sqrt{\rho}$ are two fixed or
stationary points.

First, in the case of $\rho=0$ corresponding to the Brans-Dicke
gravity or low energy effective string theory, the solution can
easily be obtained
\begin{eqnarray}
W_0 (\phi) = W_0 (\phi_i) \exp[\pm \int^{\phi}_{\phi_i} \alpha
d\phi], \label{homosol}
\end{eqnarray}
where $W_0 (\phi_i)$ is an initial value at $\phi_i$. When
$\alpha$ is non-square integrable, the solution (\ref{homosol})
for the upper $(+)$ sign grows to $\pm \infty$ as $\phi$ goes to
$\infty$ depending on the sign of $W_0 (\phi_i)$. This corresponds
to the downward motions in Fig. 1. Whereas, for the lower $(-)$
sign, the solution approaches an attractor 0 regardless of $W_0
(\phi_i)$ as $\phi$ goes to $\infty$. The solution approaches the
attractor 0 regardless of $W_0 (\phi_i)$ for the upper sign, but
it diverges to $\pm \infty$ for the lower sign depending on the
sign of $W_0 (\phi_i)$ as $\phi$ goes to $- \infty$. On the other
hand, for a square integrable $\alpha$, as $\phi$ goes to $\pm
\infty$, the solution (\ref{homosol}) approaches finite values,
$W_0 (\phi_i) \exp[\pm \int^{\pm \infty}_{\phi_i} \alpha d\phi]$,
not necessarily attractors, which correspond to the upward motions
in Fig. 1.

\begin{figure}
\includegraphics[width=0.55\linewidth,height=0.25\textheight ]{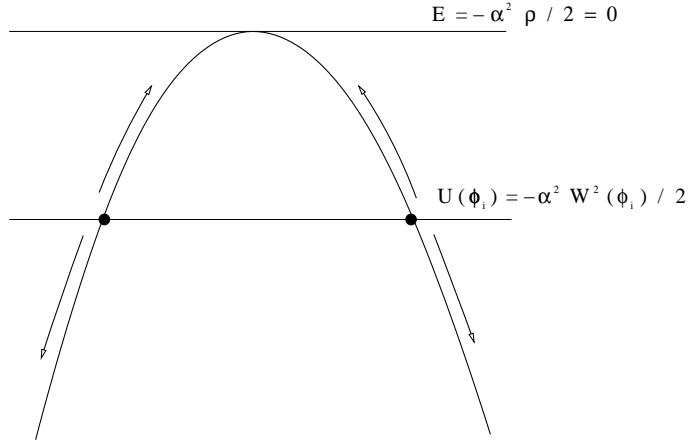}
\caption{The inverted oscillator for $\rho = 0$: $\frac{1}{2}
\Biggl(\frac{\partial W}{\partial \phi} \Biggr)^2 -
\frac{1}{2}\alpha^2 W^2 = 0$.}
\end{figure}

For general potentials with $\rho \neq 0$, we have a similar
picture as shown in Fig. 2. Not only the energy but also the
curvature of the potential depend on $\phi$. In the analogy of an
oscillator, as shown in Fig. 2,  the total energy line moves up or
down depending on $\alpha^2 \rho$ and the $\phi$-dependent
curvature $\alpha^2$ narrows or widens the parabola. As for the
$\rho=0$ case, $W$ either approaches to attractors or diverges to
$\pm \infty$, depending on the sign of $W_0 (\phi_i)$ and the
behavior of $\alpha$ and $\rho$, as $|\phi|$ grows. This behavior
of $W$ for general potentials can be calculated numerically
\cite{koh05}.

\begin{figure}
\includegraphics[width=0.55\linewidth,height=0.25\textheight ]{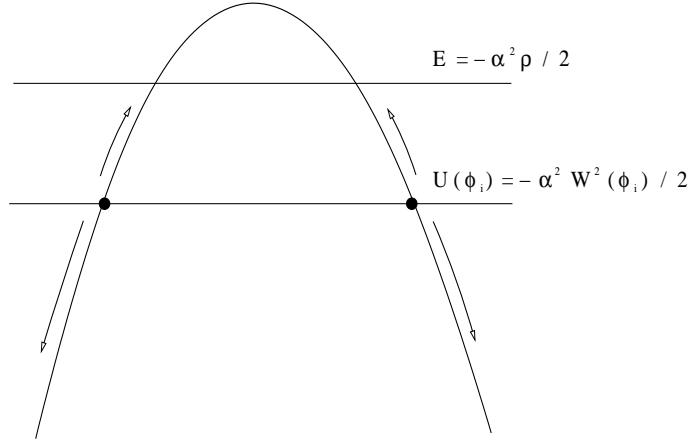}
\caption{The inverted oscillator for $\rho \neq 0$: $\frac{1}{2}
\Biggl(\frac{\partial W}{\partial \phi} \Biggr)^2 -
\frac{1}{2}\alpha^2 W^2 = - \frac{1}{2} \alpha^2 \rho$.}
\end{figure}

To find approximately an analytical solution for general
potentials, we can write $W(\phi)$ as $W(\phi)=W_0(\phi)Z(\phi)$,
where $W_0(\phi)$ is a homogeneous solution (\ref{homosol}). Then
the Hamilton-Jacobi equation reduces to the equation for
$Z(\phi)$:
\begin{eqnarray}
\Biggl(\frac{\partial Z}{\partial \phi}\Biggr)^2+2\frac{\partial
\ln W_0}{\partial \phi} Z \frac{\partial Z}{\partial
\phi}+\frac{\alpha^2 \rho}{W_0^2} =0, \label{nleqg}
\end{eqnarray}
where we have used Eq. (\ref{hjoscil}) for $\rho=0$. Equation
(\ref{nleqg}) is a nonlinear equation for $Z(\phi)$. Assuming that
$Z(\phi)$ is a slowly varying function of $\phi$, we can introduce
a small parameter $\delta$ to indicate smallness of the nonlinear
terms and rewrite Eq. (\ref{nleqg}) as
\begin{eqnarray}
2\frac{\partial \ln W_0}{\partial \phi} Z \frac{\partial
Z}{\partial \phi} +\delta \Biggl(\frac{\partial Z}{\partial
\phi}\Biggr)^2 + \frac{\alpha^2 \rho}{W_0^2} =0. \label{sveq}
\end{eqnarray}
The parameter $\delta$ will be set to one in the final result. Now
we expand $Z(\phi)$ in a series of $\delta$
\begin{eqnarray}
Z(\phi) = Z^{(0)}+\delta Z^{(1)}+ \delta^2 Z^{(2)} + \cdots.
\label{zseries}
\end{eqnarray}
Substituting Eq. (\ref{zseries}) into Eq. (\ref{sveq}) and
comparing terms of the same powers of $\delta$, we obtain the
following equations up to first order
\begin{itemize}
\item $\delta^0$:
\begin{eqnarray}
2\frac{\partial \ln W_0}{\partial \phi}Z^{(0)}\frac{\partial
Z^{(0)}}{\partial \phi} +\frac{\alpha^2 \rho}{W_0^2} =0,
\end{eqnarray}
\item $\delta^1$:
\begin{eqnarray}
2\frac{\partial \ln W_0}{\partial
\phi}\Biggl(Z^{(0)}\frac{\partial Z^{(1)}}{\partial
\phi}+Z^{(1)}\frac{Z^{(0)}}{\partial \phi}\Biggr)
+\Biggl(\frac{\partial Z^{(0)}}{\partial \phi}\Biggr)^2 =0.
\end{eqnarray}
\end{itemize}
The solutions at each order can easily be found
\begin{eqnarray}
Z^{(0)} &=& \pm \Biggl(-\int\frac{\alpha^2 \rho}{W_0^2 \partial
\ln W_0/\partial \phi}d\phi \Biggr)^{1/2}, \nonumber\\
 Z^{(1)} &=& - \frac{1}{2 Z^{(0)}} \int
 \frac{(\partial Z^{(0)}/\partial \phi)^2}{\partial W_0 /\partial
 \phi}d \phi.
\end{eqnarray}

On the other hand, for a rapidly varying function $Z(\phi)$, we
can repeat the same analysis and write Eq. (\ref{nleqg}) as
\begin{eqnarray}
\Biggl(\frac{\partial Z}{\partial \phi}\Biggr)^2+2\delta Z
\frac{\partial \ln W_0}{\partial \phi} \frac{\partial Z}{\partial
\phi}+\frac{\alpha^2 \rho}{W_0^2} =0 \label{rveq}
\end{eqnarray}
With the series expansion (\ref{zseries}) for $Z(\phi)$, we obtain
the following equations up to first order
\begin{itemize}
\item $\delta^0$:
\begin{eqnarray}
\Biggl(\frac{\partial Z^{(0)}}{\partial
\phi}\Biggr)^2+\frac{\alpha^2 \rho}{W_0^2} =0
\end{eqnarray}
\item $\delta^1$:
\begin{eqnarray}
\frac{\partial Z^{(1)}}{\partial \phi}+\frac{\partial \ln
W_0}{\partial \phi}Z^{(0)} =0
\end{eqnarray}
\end{itemize}
These equations can be simply integrated to give
\begin{eqnarray}
Z^{(0)} &=& \pm \int \sqrt{-\frac{\alpha^2 \rho}{W_0^2}}d\phi,
\nonumber\\ Z^{(1)} &=& -\int \frac{\partial \ln W_0}{\partial
\phi}Z^{(0)} d\phi.
\end{eqnarray}
The validity of the approximation in Eqs. (\ref{sveq}) and
(\ref{rveq}) for specific models will be discussed in a future
work \cite{koh05}.

\subsection{conserved quantities}

In Eq. (\ref{gen0}),  $W$ depends on $\phi$, $\gamma_{ij}$, and
$\theta$, where $\theta$ is a new canonical variable that has the
conjugate momentum
\begin{eqnarray}
\pi^{\theta}=-2\beta\gamma^{1/2}f^{3/2}(\phi) \frac{\partial
W}{\partial \theta}. \label{conmom}
\end{eqnarray}
Then the momentum constraint equation (\ref{momentum}) reduces to
\cite{salopek91}
\begin{eqnarray}
\tilde{\mathcal{H}}_i =\pi^{\theta}\partial_i \theta.
\end{eqnarray}
Since the new variable is chosen to make the new Hamiltonian
density vanish, the new Hamiltonian contains only a contribution
from the momentum constraint
\begin{eqnarray}
\tilde{H} = \int d^3 x N^i \pi^{\theta} \partial_i \theta.
\end{eqnarray}
The Hamilton equations for the new variable are
\begin{eqnarray}
\dot{\theta} -N^i\partial_i\theta   =0,
\quad \dot{\pi}^{\theta}-\partial_i(N^i\pi^{\theta})
=0.
\end{eqnarray}
In general, the new canonical variable needs not to be constant in
time, but if the spacelike hypersurfaces are chosen such that the
shift vector, $N^i$, vanishes, then $\theta$ and $\pi^{\theta}$
are constants for fixed spatial coordinates \cite{salopek91}.

It is known that a gauge invariant quantity can be derived by
taking the spatial gradient of an inhomogeneous quantity
\cite{ellis89,rigopoulos03}. By taking the spatial gradient of the
logarithm of the canonical new variable (\ref{conmom}), we can
obtain the gauge invariant quantity
\begin{eqnarray}
\partial_i \ln |\pi^{\theta}|=\frac{1}{2}\partial_i \ln \gamma
+\partial_i \phi \Biggl[\frac{3}{2}\frac{\partial \ln f}{\partial
\phi} +\frac{\partial}{\partial \phi}\ln \frac{\partial
W}{\partial \theta} \Biggr]. \label{grad}
\end{eqnarray}
To calculate the last term in the above equation, we
differentiated the Hamilton-Jacobi equation (\ref{hj}) with
respect to $\theta$:
\begin{eqnarray}
\frac{\partial}{\partial \phi}\ln \frac{\partial W}{\partial
\theta} =3\beta f^{1/2} W \frac{N}{\dot{\phi}}.
\end{eqnarray}
With this relation, Eq. (\ref{grad}) now becomes
\begin{eqnarray}
\partial_i \ln |\pi^{\theta}| =\frac{1}{2}\partial_i \ln \gamma
+\frac{NK}{\dot{\phi}}\partial_i \phi,
\end{eqnarray}
where we have used Eqs. (\ref{extrinsic}) and (\ref{scalar}). This
quantity is similar to the gauge invariant quantity $\zeta \equiv
\Psi - (3H/\dot{\phi}) \delta \phi$ in the linear perturbation
theory, which is conserved at the superhorizon scale. Here $\Psi$
is a Newtonian gravitational potential. We define the generalized
gauge invariant quantity $\zeta_i$ in nonlinear theory, which is
conserved in the large scale limit
\begin{eqnarray}
\zeta_i \equiv \frac{1}{3}\partial_i \ln |\pi^{\theta}|
=\frac{1}{6}\partial_i \ln \gamma+\frac{NK}{3\dot{\phi}}\partial_i
\phi. \label{conserve}
\end{eqnarray}
Although $\zeta_i$ is defined on the uniform energy density
hypersurfaces in general, it  coincides with a curvature
perturbation $\mathcal{R}_i$, which is defined on the comoving
hypersurfaces in a scalar field dominated universe and also
conserved in the large scale limit. $\zeta_i$ and $\mathcal{R}_i$
are proved to be conserved in the large scale limit in Refs.
\cite{rigopoulos03, lyth04}.

\section{Non-Gaussianity in generalized gravity}

Non-Gaussianity in CMB might be generated by a non-vacuum initial
state \cite{gangui02} or a nonlinear perturbation
\cite{acquaviva03}, even though the initial perturbation is
Gaussian. Although the non-vacuum initial state gives the zero
three-point correlation function, the relation between the
four-point and the two-point correlation functions, which obey the
Gaussian statistics,
\begin{eqnarray}
\langle \Biggl(\frac{\delta T}{T}\Biggr)^4 \rangle =3\langle
\Biggl(\frac{\delta T}{T}\Biggr)^2\rangle^2,
\end{eqnarray}
is no longer satisfied \cite{gangui02}. Whereas the nonlinear
gravitational perturbation leads to a non-zero three-point
correlation function. In this paper we shall focus on the
non-Gaussian signal in CMB only from nonlinear perturbations. To
show the non-Gaussianity by the nonlinear perturbation, the
gravitational potential $\Phi$ may be decomposed into a linear
part and a nonlinear part with a nonlinear parameter $f_{NL}$
\cite{komatsu01}
\begin{eqnarray}
\Phi ({\bf x}) = \Phi_{L}({\bf x})
 +f_{NL}[\Phi_{L}^2({\bf x})-\langle
\Phi_L^2({\bf x}) \rangle]. \label{nongauss}
\end{eqnarray}
Here, $\Phi_L({\bf x})$ is a linear Gaussian perturbation that has
the zero expectation value $\langle \Phi_L\rangle=0$.

With this definition, the non-vanishing component of the
$\Phi(\bf{k})$-bispectrum, which is the Fourier transform of the
three-point correlation function in the coordinate space, is
\cite{komatsu01}
\begin{eqnarray}
\langle \Phi_{L}({\bf k}_1)\Phi_{L}({\bf k}_2)\Phi_{NL}({\bf
k}_3)\rangle =2(2\pi)^3\delta^{(3)}({\bf k}_1+{\bf k}_2+{\bf k}_3)
f_{NL}P_{\Phi}({\bf k}_1)P_{\Phi}({\bf k}_2),
\end{eqnarray}
where $P_{\Phi}({\bf k})$ is the linear power spectrum given by
\begin{eqnarray}
\langle \Phi_{L}({\bf k}_1)\Phi_{L}({\bf k}_2)\rangle= (2\pi)^3
P_{\Phi}({\bf k}_1)\delta^{(3)}({\bf k}_1+{\bf k}_2),
\end{eqnarray}
and
\begin{eqnarray}
\Phi_{NL}({\bf k}) = f_{NL} \Biggl[ \int\frac{d^3
k^{\prime}}{(2\pi)^3}\Phi_{L}({\bf k}+{\bf k}^{\prime})
\Phi_{L}^{\ast}({\bf k}^{\prime})-(2\pi)^3\delta^{(3)}({\bf k})
\langle \Phi_{L}^2({\bf x})\rangle \Biggr].
\end{eqnarray}
It is known that $f_{NL}$ should be larger than order unity to be
detectable by CMB experiment. But the single field inflationary
model gives too much small value of $f_{NL} \propto
\mathcal{O}(\epsilon, \eta)$ where $\epsilon$ and $\eta$ are
slow-roll parameters \cite{acquaviva03}. Thus, if the
non-Gaussianity is detected, it can constrain inflationary models.
However, it would be interesting to calculate non-Gaussianity in
generalized gravity theories.

We follow the method in Ref. \cite{salopek90} to calculate the
nonlinear curvature perturbations on the comoving hypersurfaces.
The 3-spatial metric $\gamma_{ij}$ can be written as
\begin{eqnarray}
\gamma_{ij}(t,{\bf x}) = a^2 (t,{\bf x}) h_{ij}({\bf x}),
\end{eqnarray}
where $a(t,{\bf x})$ is a local expansion factor, and the
conformal 3-metric, $h_{ij}({\bf x})$, is independent of time and
has the unit determinant, ${\rm det} (h_{ij}) =1$. Then the local
Hubble parameter $H$  takes the form
\begin{eqnarray}
H \equiv \frac{1}{N}\frac{\dot{a}}{a} = -\frac{1}{3}K.
\end{eqnarray}
If we choose $\phi$ as a time coordinate on the comoving
hypersurfaces, we can obtain $\ln a(\phi,{\bf x})$ from Eq.
(\ref{extrinsic})
\begin{eqnarray}
\ln a(\phi,{\bf x})-\ln a(\phi_0, {\bf x}) = -\int^{\phi}_{\phi_0}
d\phi N H.
\end{eqnarray}
We use the notation $(\delta \phi)_{\mathcal{R}}(\ln a_0, {\bf x})
\equiv \partial_i \phi(\ln a_{0}, {\bf x})=\phi(\ln a_0,{\bf
x})-\phi(\ln a_0, {\bf x}_0)$ for the scalar field fluctuation on
the spatially flat hypersurfaces $(\partial_i \ln a =0)$ and
$\mathcal{R}_{\delta \phi}(\phi_0,{\bf x}) \equiv \partial_i \ln
a(\phi_0, {\bf x})=\ln a(\phi_0,{\bf x})-\ln a(\phi_0, {\bf x}_0)$ 
for the curvature perturbation on the comoving
hypersurfaces $(\partial_i \phi=0)$. The scalar field fluctuations
on the spatially flat hypersurfaces are needed to transform to the
curvature perturbation on the comoving hypersurface.  Then the
 curvature perturbation on comoving
hypersurfaces is given by
\begin{eqnarray}
\mathcal{R}_{\delta \phi} = -\int^{\phi_0 + (\delta
\phi)_{\mathcal{R}}}_{\phi_0}d\phi NH.
\end{eqnarray}
Using Eqs. (\ref{extrinsic}) and (\ref{scalar}), we get
$\mathcal{R}_{\delta \phi}$
\begin{eqnarray}
\mathcal{R}_{\delta \phi}=\frac{1}{2} \int^{\phi_0 + (\delta
\phi)_{\mathcal{R}}}_{\phi_0}d\phi \Biggl[ D \Biggl(
\frac{\partial \ln W} {\partial \phi} \Biggr)^{-1}+\frac{\partial
\ln f}{\partial \phi}\Biggr],
\end{eqnarray}
where
\begin{eqnarray}
D(\phi) = \frac{\omega(1+3\Omega)}{f}.
\end{eqnarray}
By expanding the nonlinear relation between $\mathcal{R}_{\delta
\phi}$ and $(\delta \phi)_{\mathcal{R}}$ up to second order \cite{komatsu02}, 
we obtain a nonlinear curvature perturbation, $\mathcal{R}_{\delta
\phi} = \mathcal{R}_L+\mathcal{R}_{NL}$, where
\begin{eqnarray}
\mathcal{R}_L &=& \frac{1}{2}\Biggl[ D \Biggl(\frac{\partial \ln
W} {\partial \phi} \Biggr)^{-1}+\frac{\partial \ln f}{\partial
\phi} \Biggr]
(\delta \phi)_{\mathcal{R}} , \\
\mathcal{R}_{NL} &=& \frac{1}{G} \Biggl[ \frac{\partial
D}{\partial \phi}\frac{\partial \ln W}{\partial \phi}
-D\frac{\partial^2 \ln W}{\partial \phi^2} +\Biggl(\frac{\partial
W}{\partial \phi} \Biggr)^2 \frac{\partial^2 \ln f}{\partial
\phi^2} \Biggr] (\mathcal{R}_L)^2
\end{eqnarray}
where
\begin{eqnarray}
G(\phi) = \Biggl(D+\frac{\partial \ln f}{\partial \phi}
\frac{\partial \ln W}{\partial \phi} \Biggr)^2.
\end{eqnarray}
Finally, from Eq. (\ref{nongauss}), we obtain the nonlinear
parameter
\begin{eqnarray}
f_{NL} = \frac{3}{5 G}\Biggl[ \frac{\partial D}{\partial
\phi}\frac{\partial \ln W}{\partial \phi} -D\frac{\partial^2 \ln
W}{\partial \phi^2} +\Biggl(\frac{\partial W}{\partial
\phi}\Biggr)^2 \frac{\partial^2 \ln f}{\partial \phi^2}\Biggr],
\label{nlpar}
\end{eqnarray}
where we have used the relation between $\Phi$ and $\mathcal{R}$
as $\Phi=\frac{3}{5}\mathcal{R}$  during the matter dominated era.
If we define the slow-roll parameters in generalized gravity
theories by
\begin{eqnarray}
\epsilon &=& \frac{\dot{H}}{NH^2} = \frac{\dot{\phi}}{N}
\frac{1}{H^2}\frac{\partial H}{\partial \phi}, \\
\eta &=& \frac{1}{N} \Biggl(\frac{\dot{\phi}}{N}\Biggr)^{\cdot}
\frac{N}{H\dot{\phi}},
\end{eqnarray}
using again Eqs. (\ref{extrinsic}) and (\ref{scalar}), we obtain
\begin{eqnarray}
\epsilon &=& -\frac{1}{G}\frac{\partial \ln W}{\partial \phi}
\Biggl[ D\frac{\partial \ln f}{\partial \phi} +2\Biggl\{
D+\frac{\partial^2 \ln f}{\partial \phi^2}+\frac{1}{2}
\Biggl(\frac{\partial \ln f}{\partial \phi}\Biggr)^2
-\frac{\partial \ln D}{\partial \phi}\frac{\partial \ln
f}{\partial \phi}
\Biggr\}\frac{\partial \ln W}{\partial \phi} \nonumber \\
& & + 2\frac{\partial \ln f}{\partial \phi}\frac{1}{W}
\frac{\partial^2 W}{\partial \phi^2} \Biggr],     \\
\eta &=&-\frac{1}{G}\frac{\partial \ln W}{\partial \phi} \Biggl[
D\Biggl(\frac{\partial \ln f}{\partial \phi}-2\frac{\partial \ln
D} {\partial \phi}\Biggr)+ 2D\Biggl(\frac{\partial W}{\partial
\phi}\Biggr)^{-1}
\frac{\partial^2 W}{\partial \phi^2} \nonumber \\
& &+\Biggl\{\Biggl(\frac{\partial \ln f} {\partial
\phi}\Biggr)^2-2\frac{\partial \ln f}{\partial \phi}
\frac{\partial \ln D}{\partial \phi}\Biggr\}\frac{\partial \ln
W}{\partial \phi} +2\frac{\partial \ln f}{\partial
\phi}\frac{1}{W}\frac{\partial^2 W} {\partial \phi^2} \Biggr].
\end{eqnarray}
The nonlinear part of the curvature perturbation can be written in
terms of the slow-roll parameters, $\epsilon$ and $\eta$, as
\begin{eqnarray}
\mathcal{R}_{NL}=\frac{1}{2}(\eta-\epsilon)(\mathcal{R}_L)^2.
\end{eqnarray}
Hence, the nonlinear parameter, $f_{NL}$, for generalized gravity
theories takes the form
\begin{eqnarray}
f_{NL} = \frac{1}{2}(\eta-\epsilon). \label{unv non}
\end{eqnarray}
Note that the nonlinear parameter (\ref{unv non}) in generalized
gravities has the same form as in Einstein gravity. The WMAP
result gives a constraint on $f_{NL}$ by $-58 <f_{NL}<134$ at
$95\%$ CL \cite{komatsu03}. Since the slow-roll inflation implies
that $|\epsilon|, |\eta| \ll 1$, non-Gaussianity in the single
scalar field inflation model is difficult to be observed by CMB
experiments.

\section{Conclusion}

It is expected that nonlinear perturbations may be responsible for
a non-Gaussian signal in CMB experiments. The Hamilton-Jacobi
theory can provide a useful and convenient tool to deal with
nonlinear perturbations. In this paper we derived the
Hamilton-Jacobi equation in generalized gravity theory. Through a
canonical transformation, a new set of canonical variables was
chosen that could make the new Hamiltonian density vanish. The
conserved quantity, $\zeta_i =
\partial_i \ln a- (NH/\dot{\phi})\partial_i \phi$, was obtained
in the large scale limit by using the fact that the new canonical
variable is constant in time on fixed spacelike hypersurfaces. The
$\zeta_i$ could be regarded as a generalization of the gauge
invariant quantity $\zeta = \Psi - (3H/\dot{\phi})\delta \phi$ in
the linear perturbation theory. In the long wavelength
approximation, we found the exact solutions to the lowest order
Hamilton-Jacobi equation for special scalar potentials and
introduced an approximation scheme for general potentials.

The non-Gaussianity from nonlinear density perturbations was
parameterized with a nonlinear parameter $f_{NL}$, which is an
expansion parameter of the gravitational potential in Eq.
(\ref{nongauss}). The nonlinear parameter can be measured by CMB
observations. Any detection of the non-Gaussianity of CMB may put
a strong constraint on inflation models. Especially, $f_{NL}$
predicted by the single field slow-roll inflation in Einstein
gravity is too small to be detected in CMB experiments through a
non-Gaussian signal. Nevertheless, it would be interesting to
investigate the non-Gaussianity in generalized gravity theory. We
have found that even in generalized gravity theory, the nonlinear
parameter for the slow-roll inflation takes the same form $f_{NL}
= (\eta - \epsilon)/2$ as in Einstein gravity. Hence, the
slow-roll inflation models based on Brans-Dicke theory,
non-minimally coupled scalar field theory, and low energy
effective string theory, {\it et al} as well as Einstein gravity
will have the same order of non-Gaussianity for CMB. This has a
physical implication that slow-roll inflation models with a single
scalar field in such alternative theories of gravity will be ruled
out if a signal for non-Gaussianity is observed in future CMB
experiments. Therefore, the non-Gaussianity would require
multi-field inflation models or a different generating mechanism
for density perturbations \cite{bartolo04} in generalized gravity.

\acknowledgments This work was supported by Korea Astronomy
Observatory (KAO).


\end{document}